\documentstyle[preprint,aps,version2]{revtex}
\begin{document}

\begin{title}
{Asymptotically exact solution of a local copper-oxide model}
\end{title}

\author{Guang-Ming Zhang and Lu Yu\cite{yu}}
\begin{instit}
{International Center for Theoretical Physics, P. O. Box 586, 34100
 Trieste, Italy.}
\end{instit}

\begin{abstract}
We present an asymptotically exact solution of a local copper-oxide model
abstracted from the multi-band models. The phase diagram is obtained
through the renormalization-group analysis of the partition function. In the
strong coupling regime, we find an exactly solved line, which crosses the
quantum critical point of the mixed valence regime separating two different
Fermi-liquid (FL) phases. At this critical point, a many-particle
resonance is formed near the chemical potential, and a marginal-FL spectrum
can be derived for the spin and charge susceptibilities.
\end{abstract}


\narrowtext
\newpage
One of the most challenging theoretical issues has been raised by the
experimental observations on copper-oxide based metals:
What is the appropriate description of the normal state which does not
fit into the Fermi-liquid (FL) phenomenology. In an attempt to
find the unifying features in the diverse observed anomalies, a
phenomenological marginal-FL (MFL) spectrum for the spin and charge
fluctuations was proposed by Varma {\it et al.} [1]. The essential
point is that the frequency dependence of the susceptibilities is singular,
while the momentum dependence is assumed smooth. It is expected that the
study of the related impurity models would shed some light on the basic
physics of the lattice models, in the same sense as the Anderson model vs
the single band Hubbard model.

Recently, Varma {\it et al.} [2,3] have explored the local
non-FL properties of a generalized Anderson model, equivalent to a
single-impurity version of one of the multi-band models proposed to describe
the physics of the copper-oxide compounds [4,5].
In fact, such a model was first studied by Haldane in the late 70s on
Anderson's suggestion [6]. They realized that the usual Anderson
model should be supplemented by screening channels to saturate the Friedel
sum rule, which is crucial for the understanding of the mixed valence physics.
The recent Wilson's renormalization-group (RG) study [2] and
analytical arguments [3] have provided a microscopic scenario for
the MFL phenomenology within this generalized Anderson model.

In this Letter, we derive explicitly physical properties of this
local copper-oxide model through exactly solving the Hamiltonian in
the asymptotic limit. We first separate the model into "hybridizing" and
"screening" parts. The first part is the usual Anderson model plus
X-ray edge-like (XRE) scattering terms, which were neglected in all previous
studies [2,3,6]. This part reduces to a
hybridization process and potential scatterings, which can be
merged into
an effective hybridization via a canonical transformation, giving rise to
a local FL behavior. The second part is the multi-channel XRE scattering
terms, and can be reduced to a single spinless channel, exhibiting the
Anderson catastrophe. These XRE singularities can also be transformed into
the effective
hybridization, so the physics of the full model results from a
competition between the above two different physical factors in the
hybridization. Then we apply the RG theory [7] to derive flow equations
from which a phase diagram is found. There exist two different
FL phases: a Kondo phase and a local free FL phase separated by a mixed
valence phase, displaying a local MFL behavior. Finally, from the
partition function and the effective Hamiltonian, we take the strong coupling
limit, and an exactly solved line (analogue of the Toulouse limit in the
Kondo problem) can be found, which crosses the quantum critical
point of the mixed valence phase where a many-particle
resonance is formed between the localized impurity and the conduction
electrons. Properties of the above three phases, especially, the mixed
valence phase controlled by the critical point can be exactly derived.

The local copper-oxide model is specified by the Hamiltonian [2]
\widetext
\begin{eqnarray}\label{H}
&& H=H_h+H_s, \nonumber \\ &&
H_h=\sum_{k,\sigma}\epsilon_{k}C^{\dag}_{k,\sigma,0}C_{k,\sigma,0}
+\epsilon_{d}n_d+Un_{d,\uparrow}n_{d,\downarrow}+
\frac{t}{\sqrt{N}}\sum_{k,\sigma}(C^{\dag}_{k,\sigma,0}d_{\sigma}+h.c.)
\nonumber \\ && \hspace{1cm}
+\sum_{k,k',\sigma}\frac{V_0}{N}C^{\dag}_{k,\sigma,0}C_{k',\sigma,0}
(n_{d,\sigma}-\frac{1}{2})
+\sum_{k,k',\sigma}\frac{V'_0}{N}C^{\dag}_{k,\sigma,0}C_{k',\sigma,0}
(n_{d,{\bar \sigma}}-\frac{1}{2}),\nonumber \\ &&
 H_s=\sum_{k,\sigma,l>0}\epsilon_{k}C^{\dag}_{k,\sigma,l}C_{k,\sigma,l}
+\sum_{k,k',\sigma,l>0}\frac{V_l}{N}C^{\dag}_{k,\sigma,l}C_{k',\sigma,l}
(n_d-\frac{1}{2}),
\end{eqnarray}
where $n_d=\sum_{\sigma}n_{d,\sigma}$, $N$ is the number of the lattice
sites, and we have separated
the Hamiltonian into hybridizing and screening parts and distinguished
parallel-spin and opposite-spin XRE scatterings in the hybridizing
channel. The localized impurity hybridizes only with channel $l=0$. The
chemical potential $\mu$ is set to zero, and we are interested in the
case when the local impurity level $\epsilon_d$ is close to zero. The
spinless version of this
model, {\it i.e.} the multi-channel resonant level model, has been solved
exactly [8,9], displaying a FL vs non-FL transition as the interaction
parameters are varied. In fact, $H_h$ is the usual Anderson model
plus XRE potential scatterings, while $H_s$ is the multi-channel XRE
Hamiltonian. It is more convenient to take an infinite $U$ limit as in the
usual treatments for the Anderson model (the low-energy physics is kept).
Thus, we add a local constraint for the local impurity $n_d\leq 1$.

First, we use abelian bosonization to handle the XRE singularities of
the screening channels, which reduce to a one-dimensional problem with
only one Fermi point for each channel and the dispersion is
linearized with a cutoff $k_D$ [10]. Since the spin degrees of freedom of
the screening channel electrons are trivially involved, they can be
separated from $H_s$. Moreover, we can assume $V_l=V_s$ for all $l>0$
without loss of generality, so the channel index can be dropped. Thus,
all the screening channels are described by a single spinless channel [3].
The resulting form is
\begin{equation}\label{H_s}
H_s^{b}=\sum_{k>0}\frac{k}{\rho}a^{\dag}_{k}a_{k}
+\sum_{k>0}{\tilde{V}_s}\sqrt{\frac{k}{N}}(a^{\dag}_{k}+
a_{k})(n_d-\frac{1}{2}),
\end{equation}
with $\tilde{V}_s\equiv\sqrt{2N_s}V_s$ and $N_s$ is the number of the
screening channels. $a^{\dag}_{k}$ and $a_{k}$ are the bosonic operators
describing the charge degrees of freedom of the screening channel and
$\rho=(hv_F)^{-1}$ is the density of states at the Fermi point. Employing
the inverse bosonization, we can transform bosons back to fermions:
$H_s^{f}=\sum_{k}\epsilon_{k}s^{\dag}_{k}s_{k}
+\frac{\tilde{V}_s}{N}\sum_{k,k'}s^{\dag}_{k}s_{k'}(n_d-\frac{1}{2})$.
The Hamiltonian (\ref{H_s}) can be diagonalized through a canonical
transformation [8]:
$ U=exp\{\sum_{k>0}\frac{\delta_s/\pi}{\sqrt{kN}}
(a_k-a_k^{\dag})(n_d-\frac{1}{2})\}$,
where $\delta_s\approx \pi\rho\tilde{V}_s$ is the phase shift generated
by the XRE scattering of all screening channels at
the impurity.

For the hybridizing channel, employing another canonical transformation
[8,11]:
$ S= exp\{\sum_{\{k>0,\sigma\}} \frac{\delta_0/\pi}{\sqrt{kN}}
(b_{k,\sigma} - b_{k,\sigma}^{\dag}) (n_{d,\sigma} - \frac{1}{2})\}
exp\{\sum_{\{k>0,\sigma\}} \frac{\delta'_0/\pi}{\sqrt{kN}}
(b_{k,\sigma} - b_{k,\sigma}^{\dag}) (n_{d,{\bar \sigma}} - \frac{1}{2})\},$
where $\delta_0\approx\pi\rho V_0$ and $\delta '_0\approx\pi\rho V'_0$
are the phase shifts induced by the parallel-spin ($\sigma$) and
opposite-spin ($\bar\sigma$)
XRE scattering of the hybridizing channel, respectively, we merge XRE
scatterings into the hybridization and obtain
\begin{equation}
 H_{eff} =\sum_{k>0,\sigma}\frac{k}{\rho}b^{\dag}_{k,\sigma}b_{k,\sigma}
  +\epsilon'_{d}n_d
  +t\sum_{\sigma}(\Delta\psi^{\dag}_{\sigma}d_{\sigma}+h.c.)
+\sum_{k>0}\frac{k}{\rho}a^{\dag}_{k}a_{k},
\end{equation}
with $\psi_{\sigma}\equiv\sqrt{k_D}
exp[\sum_{k>0}\frac{1-\delta_0/\pi}{\sqrt{kN}}(b_{k,\sigma} -
b_{k,\sigma}^{\dag})]
exp[-\sum_{k>0} \frac{\delta'_0/\pi}{\sqrt{kN}}(b_{k,{\bar \sigma}
} - b_{k,{\bar \sigma}}^{\dag})]$,
$\Delta\equiv exp{[\sum_{k>0}\frac{\delta_s/\pi}{\sqrt{kN}}
(a_k-a_k^{\dag})]}$, $\epsilon'_{d}\equiv\epsilon_d-\rho V_0 V'_0$, and
$n_d\leq 1$. A fermionic form is obtained by inverse bosonization
\begin{equation}\label{H_eff}
\tilde{H}_{eff}=\sum_{k,\sigma}\epsilon_k C^{\dag}_{k,\sigma}C_{k,\sigma}
+\epsilon'_{d}n_d+t\sum_{\sigma}(\Delta\psi^{\dag}_{\sigma}d_{\sigma}+h.c.)
+\sum_{k}\epsilon_{k}s^{\dag}_{k}s_{k},
\end{equation}
where $\psi_{\sigma}=(k_D)^{\frac{\delta_0}{2\pi}+\frac{\delta '_0}{2\pi}}
(C_{\sigma})^{(1-\frac{\delta_0}{\pi})}(C_{{\bar \sigma}})^{-\frac{\delta
'_0}{\pi}}$, $C_{\sigma}=\frac{1}{\sqrt{N}}\sum_k C_{k,\sigma}$,
$\Delta=(k_D)^{-\frac{\delta_s}{2\pi}}(s_0)^{\frac{\delta_s}{\pi}}$,
$s_0=\frac{1}{\sqrt{N}}\sum_k s_k$, and $n_d\leq 1$.

Next, we derive the partition function, dividing $\tilde{H}_{eff}$ as
$\tilde{H}_{eff}=\tilde{H}_0+\tilde{H}_I$ with $\tilde{H}_0$ and
$\tilde{H}_I$ as
the free and the hybridization parts of (\ref{H_eff}), respectively.
Paralleling all the strategies of previous studies [6-8], we
write the partition function in terms of a sum over
histories of the impurity. Each history is a sequence of transitions
between the three local $d$ states $\mid\alpha>=\mid 0>$, and
$\mid\sigma>$, $\sigma=\uparrow,\downarrow$. The transitions take place at
the imaginary time $0<\tau_1<...<\tau_n<\beta=\frac{1}{T}$: along the
Feymann trajectory the local state is at $\mid\sigma_{i+1}>$ from $\tau_i$
to $\tau_{i+1}$ ($i$=1 to n). The partition function of (\ref{H_eff}) is
now given by
\widetext
\begin{eqnarray}\label{Z}
&& Z=\sum_{n=0}^{\infty}
\sum_{\{\sigma,\sigma_1,..,\sigma_{n+1}=\sigma\}}
\sum_{\{\alpha=0,\sigma\}}
(\Gamma\tau)^n
\int_{0}^{\beta}\frac{d\tau_{2n}}{\tau}
\int_0^{\tau_{2n}-\tau}\frac{d\tau_{2n-1}}{\tau}...\int_0^{\tau_2-\tau}
\frac{d\tau_1}{\tau}
\Pi_{i} y_{\sigma_i,\sigma_{i-1}}^{\alpha} \nonumber \\ &&
exp\{E_d^{\alpha}\sum_i (-1)^i \tau_i + \sum_{i>j}(-1)^{i+j}
[(1-\frac{\delta_0}{\pi})^2\delta_{\sigma_i,\sigma_j}+(\frac{\delta
'_0}{\pi})^2\delta_{\sigma_i,\sigma_j}+
(\frac{\delta_s}{\pi})^2]ln\mid\frac{\tau_i-\tau_j}{\tau}\mid\},
\end{eqnarray}
\narrowtext
\noindent
where the bare hybridization strengths are defined as $\Gamma=\rho t^2$,
while the cutoff factor $\tau=\frac{\rho}{k_D}$. The effective "magnetic
field" reflects the differences of the local state energies:
$E_d^0=-\epsilon'_d$, $E_d^{\sigma}=\epsilon'_d$. The fugacity
$(y_{\sigma_i,\sigma_{i+1}}^{\alpha}{\sqrt{\Gamma\tau}})$ is the
amplitude associated with a transition from $\mid\sigma_i>$ to
$\mid\sigma_{i+1}>$,
with $y_{\sigma_{2i},\sigma_{2i-1}}^{0}=\delta_{\sigma_{2i},\sigma_{2i-1}}$,
$y_{\sigma_{2i+1},\sigma_{2i}}^{0}=1$;
$y_{\sigma_{2i},\sigma_{2i-1}}^{\sigma}=1$,
$y_{\sigma_{2i+1},\sigma_{2i}}^{\sigma}=\delta_{\sigma_{2i+1},\sigma_{2i}}$.
The long-range logarithmic interaction between the flipping events arises
from the reaction of the conduction electron bath towards the transition
between the local states. The local disturbance on the
bath involves two factors: the absorption or
emission of the local conduction electrons and the change in the local
potential that the conduction electrons experience [8]. Both kinds of
disturbance are incorporated in the effective "charge" factor, {\it i.e.}
the coefficient of the logarithmic function of (\ref{Z}).

The partition function (\ref{Z}) could be obtained
directly from the model Hamiltonian (\ref{H}) without bosonization, using
the famous fermion techniques [12] in certain asymptotic limit which, we
believe, is also valid here. This alternative
derivation would allow us to rectify the phase shifts obtained by the
bosonization treatments to the exact expressions:
$\delta_s=2tan^{-1}(\frac{\pi}{2}\rho\tilde{V}_s)$,
$\delta_0=2tan^{-1}(\frac{\pi}{2}\rho {V}_0)$, and
$\delta '_0=2tan^{-1}(\frac{\pi}{2}\rho {V'}_0)$ so that the corresponding
effective Hamiltonian (\ref{H_eff}) might be used beyond the range of
validity for the bosonization method, especially in the following strong
coupling limit where the renormalized parameters of the model recover their
bare values.

To set up the RG flow equations, we can directly employ the
scaling theory proposed by  Anderson {\it et al.} [7] in
the Coulomb gas representation. The RG equations describe the flow
behavior as the bandwidth is reduced. They are given by
\begin{equation}
 \frac{d{\Gamma}}{dln\tau}=-\gamma{\Gamma}, \hspace{0.5cm}
 \frac{d\epsilon_d}{dln\tau}\approx\frac{\Gamma}{\pi}, \hspace{0.5cm}
 \frac{d\gamma}{dln\tau}\approx -(\gamma+1)^2\Gamma\tau,
\end{equation}
with $\gamma\equiv-\frac{2\delta_0}{\pi}+(\frac{\delta_0}{\pi})^2+
(\frac{\delta '_0}{\pi})^2+ (\frac{\delta_s}{\pi})^2$ describing the total
interaction strength between the conduction electrons and the local
impurity, which should be positive in our case.
These equations were derived by assuming $\Gamma\tau\leq 1$, a rare gas of
spin-flips. In the zeroth
order, we can construct two invariants ($\Gamma^*\tau^*=1$) :
$\Gamma^*=\Gamma(\Gamma\tau_0)^{\frac{\gamma}{1-\gamma}}$ and
$\epsilon_d^*=\epsilon_d+\frac{\Gamma}{\gamma\pi}[1-(\Gamma\tau_0)
  ^{\frac{\gamma}{1-\gamma}}]$, where $\epsilon_d$, $\Gamma$,
and $\tau_0$ are initial (bare) parameters. In terms of these
scaling invariants, the running resonance width and impurity level are
written as
$\Gamma(\tau)=(\Gamma^*)^{1-\gamma}(\tau)^{-\gamma}$ and
$\epsilon_d(\tau)=\epsilon^*_d+\frac{\Gamma^*}{\gamma\pi}
[1-(\Gamma^*\tau)^{-\gamma}]$.
Obviously, for the case $V_0=V'_0=V_s=0$, {\it i.e.} a trivial limit of
$\gamma=0$, the two expressions become
$\Gamma(\tau)=\Gamma^*$ and
$\epsilon_d(\tau)=\epsilon^*_d+\frac{\Gamma^*}{\pi}ln(\Gamma^*\tau)$, which
exactly recovers the Haldane's RG results for the standard Anderson model
[7]. Moreover, a complete phase diagram can be determined by comparing
the invariants $\epsilon^*_d$ and $\Gamma^*$. In the plane
$\Gamma^*-\gamma$ (Fig.1), there are three phases corresponding to
different impurity-occupancies: single-occupancy regime
($\epsilon^*_d\ll-\Gamma^*$) where a singlet state is formed and the Kondo
effect shows up; zero-occupancy regime ($\epsilon^*_d\gg\Gamma^*$) where the
model corresponds to a local free FL phase, and the mixed valence regime
($\mid\epsilon^*_d\mid\leq\Gamma^*$) where $<n_d>$ fluctuates between 0
and 1 phases. The mixed valence
regime separates the $< n_d >=1$ and the $<n_d>=0$ phases with crossover
lines $\Gamma^*\approx\frac{\Gamma}{1-\gamma\pi}$
and $\Gamma^*\approx\frac{\Gamma}{1+\gamma\pi}$, respectively. Although
parameter $\gamma$ is not renormalized in the zeroth order of
$\Gamma\tau$, in its first order it is renormalized to
smaller values as $\tau$ increases. On the other hand,  from
$\Gamma^*\approx-\gamma\pi\epsilon^*_d+\Gamma$, $\Gamma^*$ increases when
$\epsilon_d^*>0$ and decreases, while it decreases if $\epsilon_d^*<0$ and
decreases in absolute value. Hence the flow directions indicated in Fig.1.
In the end, the renormalized $\gamma$ tends to zero as
$\epsilon^*_d\rightarrow 0$, $\Gamma^*\rightarrow\Gamma$.
However, when $\Gamma(\tau)\tau\approx 1$, the RG equations are not
correct quantitatively.

In addition to this RG analysis, a strong coupling
limit can be extracted {\it independently} from the partition function
(5) and the effective Hamiltonian (4).
As follows from (4), when opposite-spin XRE scattering in the hybridizing
channel renormalizes to zero
while the XRE scatterings of the parallel-spin and the screening
channel reach their respective unitary limit, {\it i.e.}
$\delta_0=\delta_s=\pi$ and $\delta '_0=0$ or $\gamma=0$, the phase shifts
due to hybridization and parallel-spin XRE scattering in
the hybridizing channel compensate each other, thus the
hybridizing electrons become completely free.
 In such a strong coupling limit, the effective Hamiltonian
becomes \begin{equation}\label{H_T}
 H_T=\sum_{k,\sigma}\epsilon_k C^{\dag}_{k,\sigma}C_{k,\sigma}
+\epsilon_{d}n_d+t\sum_{\sigma}(s_0^{\dag}d_{\sigma}^{\dag}+h.c.)
+\sum_{k}\epsilon_k s^{\dag}_{k}s_{k},
\end{equation}
with constraint $n_d\leq 1$ or $n_d+s^{\dag}_0s_0=1$, which
reflects the Friedel sum rule in this limit.  It is obvious that the
partition function derived from the Hamiltonian (\ref{H_T}) is exactly the
same as (\ref{Z}) for $\delta_0=\delta_s=\pi$ and $\delta '_0=0$, {\it i.e.}
$\gamma=0$. In this sense, the strong coupling limit found here is somehow
analogous to the Toulouse limit of the Kondo problem [13],
although the actual physics involved is quite different. The most essential
difference is that the unitary limit has been actually reached in our
case. The vanishing of opposite-spin XRE scattering in the
hybridizing channel is exactly what is
required by the infinite U limit, because any finite
hybridization between opposite-spin hybridizing electrons and the local
impurity (to compensate XRE scattering) will contradict the single occupancy
constraint.

In Fig.1, $\gamma=0$ is a strong coupling limit line of this local
copper-oxide model. In (\ref{H_T}), only charge of the local
impurity $\alpha\equiv\frac{1}{\sqrt{2}}(d_{\uparrow}+d_{\downarrow})$ is
coupled to the conduction electrons, while the spin
$\beta\equiv\frac{1}{2}(d_{\uparrow}-d_{\downarrow})$ is decoupled
except for the constraint. Thus, in this limit, the Hamiltonian is:
$ \tilde{H}_T=\sum_{k}\epsilon_k s^{\dag}_{k}s_{k}+
\epsilon_{d}(n_{\alpha}+n_{\beta})+\sqrt{2}t (s_0^{\dag}\alpha^{\dag}+h.c.)$,
where $n_{\alpha}+n_{\beta}\le 1$ and  the hybridizing electrons do not
show up explicitly. This Hamiltonian is essentially the same as Eq.
(14) of [3].  Since $\tilde{H}_T$ conserves $n_{\beta}$, we can
calculate physical quantities by taking the trace on its two subspaces
$n_{\beta}=0,1$.

(i). When $\epsilon_d\gg\epsilon_{dc}$, a critical value to be
defined later, the $n_{\alpha}=n_{\beta}=0$
state is favored in the low-energy regime. All charge fluctuation processes
are frozen out and $s^{\dag}_0s_0=1$ at the impurity site so that the Friedel
sum rule is saturated. The hybridization strength ${\Gamma^*}$
should be renormalized to zero, and a local free FL behavior is thus
displayed [14].

(ii). The opposite case $\epsilon_d\ll\epsilon_{dc}$ favors
$n_{\alpha}+n_{\beta}=1$ in the low-energy regime, and there are two
possible configurations: $n_{\alpha}=0$, $n_{\beta}=1$; $n_{\alpha}=1$,
$n_{\beta}=0$. All charge fluctuation processes are also frozen out but
$s^{\dag}_0s_0=0$. The hybridization strength ${\Gamma^*}$ should be
renormalized to $+\infty$, and the system is scaled to Wilson's strong
coupling fixed point of the Kondo problem: a local FL behavior in its
unitarity limit [14]. The symmetry of the ground state for the present
case is different from that at $\epsilon_d\gg\epsilon_{dc}$, so we
anticipate a quantum critical point at $\epsilon_d=\epsilon_{dc}$.

(iii). When $\epsilon_d\rightarrow\epsilon_{dc}$, the
localized impurity fluctuates between zero- and single-occupancy
$<n_{\alpha}+n_{\beta}>\rightarrow\frac{1}{2}$, and hybridizes
with only part of the screening electron
$<s^{\dag}_0s_0>\rightarrow\frac{1}{2}$, corresponding to the mixed valence
phase.
At the special point ($\gamma=0$, ${\Gamma^*}=\Gamma$) along the strong
coupling line, the above two different FL states are degenerate.
This special point is just the quantum critical point controlling the
physics of the whole mixed valence phase. An analogous quantum critical
point was found in the two-channel or two-impurity Kondo problems
[15]. At zero temperature, we find
$\epsilon_{dc}\approx -\frac{3ln2}{\pi}\Gamma$ and
$<n_{\alpha}>\approx\frac{1}{2}$, $<n_{\beta}>\approx 0$.
Thus, the local impurity level is close to
the chemical potential. Using the phase-shift representation of the
Friedel sum rule:
$<n_d>+\frac{1}{\pi}\delta_h(\mu)+\frac{1}{\pi}\delta_s(\mu)=1$, we
easily obtain the phase shift of the screening electrons caused by the final
hybridization is $\frac{\pi}{2}$ at the chemical potential.
Since both hybridizing and screening electrons are involved and
there is a constraint on the local impurity, the hybridization
becomes a many-particle resonance, drawing some weight of the one-particle
spectra from higher energies at the scale of the charge transfer gap in
the insulating state.
Such a many-particle resonance breaks down the Landau correspondence
between the low-lying excitations of the interacting and non-interacting
fermions [3]. At finite low temperatures, the impurity charge and
longitudinal spin susceptibilities $\chi_{\sigma,\rho}$ can also be
calculated using the relations
$\sigma_z=(\alpha^{\dag}\beta+\beta^{\dag}\alpha)$,
$\rho=(\alpha^{\dag}\alpha+\beta^{\dag}\beta-\frac{1}{2})$.
It has been found that $\chi_{\sigma,\rho}$ are proportional to
$\Gamma^{-1}ln(\frac{\Gamma}{T})$, as expected from
the MFL phenomenology [3]. Thus, the MFL behavior controls the whole
mixed valence regime.

In conclusion, we have presented an asymptotically exact solution of the local
copper-oxide model including the charge fluctuations of the screening
electrons. The physical picture of the breakdown of the FL behavior
pointed out in Ref.[3] is basically correct. However, the
justification of their physical arguments involves several unclear
approximations. The crucial point is that in all the previous studies
[2,3,6], the parallel-spin XRE
scattering in the hybridizing channel was assumed to be zero, {\it i.e.}
$V_0=0$.  However, namely this assumption obscures the physical
features of the model and makes the problem much more involved. Due to the
emergence of new relevant variables in their theory, in principle, they
can not reach the strong-coupling limit. Since our solution
is based on both RG flow analysis and the strong coupling effective
Hamiltonian, the MFL behavior should be a universal
property of the mixed valence phase. Of course, whether a specific
system is at the quantum critical point depends on a special combination
of parameters. A more interesting question, whether the chemical
potential of a real mixed valence system is pinned at the critical point,
requries further studies.

{\it Acknowledgements}. G. M. Zhang is grateful to Z. B. Su and
X. Y. Zhang for many useful discussions and earlier collaboration
on a related subject. L. Yu thanks C. M. Varma for providing and
explaining the paper [3] prior to publication, as well as stimulating
discussions.

\vskip 2cm
Figure Caption
\vskip .5cm
Fig.1. The phase diagram of the local copper-oxide model. I, II, and
III are single-occupancy phase, mixed valence phase, and zero-occupancy
phase, respectively. The two thick lines correspond to the crossover
lines, while the two slender lines with arrows roughly show up the flow
directions and the horizontal one is $\epsilon_d^*=0$ line.

\newpage
\setlength{\unitlength}{0.240900pt}
\ifx\plotpoint\undefined\newsavebox{\plotpoint}\fi
\sbox{\plotpoint}{\rule[-0.175pt]{0.350pt}{0.350pt}}%
\begin{picture}(900,900)(0,0)
\tenrm
\sbox{\plotpoint}{\rule[-0.175pt]{0.350pt}{0.350pt}}%
\put(264,158){\rule[-0.175pt]{137.795pt}{0.350pt}}
\put(264,158){\rule[-0.175pt]{0.350pt}{151.526pt}}
\put(264,158){\rule[-0.175pt]{4.818pt}{0.350pt}}
\put(242,158){\makebox(0,0)[r]{0}}
\put(816,158){\rule[-0.175pt]{4.818pt}{0.350pt}}
\put(264,158){\rule[-0.175pt]{0.350pt}{4.818pt}}
\put(264,113){\makebox(0,0){0}}
\put(264,767){\rule[-0.175pt]{0.350pt}{4.818pt}}
\put(836,158){\rule[-0.175pt]{0.350pt}{4.818pt}}
\put(836,113){\makebox(0,0){1}}
\put(836,767){\rule[-0.175pt]{0.350pt}{4.818pt}}
\put(264,158){\rule[-0.175pt]{137.795pt}{0.350pt}}
\put(836,158){\rule[-0.175pt]{0.350pt}{151.526pt}}
\put(264,787){\rule[-0.175pt]{137.795pt}{0.350pt}}
\put(550,68){\makebox(0,0){$\gamma$}}
\put(150,577){\makebox(0,0)[l]{$\Gamma^*$}}
\put(207,368){\makebox(0,0)[l]{$\Gamma$}}
\put(298,577){\makebox(0,0)[l]{I}}
\put(607,525){\makebox(0,0)[l]{II}}
\put(407,221){\makebox(0,0)[l]{III}}
\put(264,158){\rule[-0.175pt]{0.350pt}{151.526pt}}
\put(510,294){\vector(0,1){2}}
\put(393,349){\rule[-0.175pt]{1.847pt}{0.350pt}}
\put(385,350){\rule[-0.175pt]{1.847pt}{0.350pt}}
\put(378,351){\rule[-0.175pt]{1.847pt}{0.350pt}}
\put(378,352){\vector(-4,1){0}}
\put(353,460){\vector(0,-1){19}}
\put(378,352){\usebox{\plotpoint}}
\sbox{\plotpoint}{\rule[-0.500pt]{1.000pt}{1.000pt}}%
\put(264,368){\usebox{\plotpoint}}
\put(264,368){\usebox{\plotpoint}}
\put(265,369){\usebox{\plotpoint}}
\put(266,370){\usebox{\plotpoint}}
\put(267,371){\usebox{\plotpoint}}
\put(268,372){\usebox{\plotpoint}}
\put(269,373){\usebox{\plotpoint}}
\put(270,374){\usebox{\plotpoint}}
\put(271,376){\usebox{\plotpoint}}
\put(272,377){\usebox{\plotpoint}}
\put(273,378){\usebox{\plotpoint}}
\put(274,379){\usebox{\plotpoint}}
\put(275,380){\usebox{\plotpoint}}
\put(276,381){\usebox{\plotpoint}}
\put(277,383){\usebox{\plotpoint}}
\put(278,385){\usebox{\plotpoint}}
\put(279,386){\usebox{\plotpoint}}
\put(280,388){\usebox{\plotpoint}}
\put(281,390){\usebox{\plotpoint}}
\put(282,391){\usebox{\plotpoint}}
\put(283,392){\usebox{\plotpoint}}
\put(284,394){\usebox{\plotpoint}}
\put(285,395){\usebox{\plotpoint}}
\put(286,396){\usebox{\plotpoint}}
\put(287,398){\usebox{\plotpoint}}
\put(288,399){\usebox{\plotpoint}}
\put(289,401){\usebox{\plotpoint}}
\put(290,402){\usebox{\plotpoint}}
\put(291,404){\usebox{\plotpoint}}
\put(292,405){\usebox{\plotpoint}}
\put(293,407){\usebox{\plotpoint}}
\put(294,408){\usebox{\plotpoint}}
\put(295,410){\usebox{\plotpoint}}
\put(296,411){\usebox{\plotpoint}}
\put(297,413){\usebox{\plotpoint}}
\put(298,415){\usebox{\plotpoint}}
\put(299,416){\usebox{\plotpoint}}
\put(300,419){\usebox{\plotpoint}}
\put(301,421){\usebox{\plotpoint}}
\put(302,423){\usebox{\plotpoint}}
\put(303,425){\usebox{\plotpoint}}
\put(304,428){\usebox{\plotpoint}}
\put(305,429){\usebox{\plotpoint}}
\put(306,431){\usebox{\plotpoint}}
\put(307,433){\usebox{\plotpoint}}
\put(308,435){\usebox{\plotpoint}}
\put(309,437){\usebox{\plotpoint}}
\put(310,439){\usebox{\plotpoint}}
\put(311,441){\usebox{\plotpoint}}
\put(312,443){\usebox{\plotpoint}}
\put(313,445){\usebox{\plotpoint}}
\put(314,447){\usebox{\plotpoint}}
\put(315,449){\usebox{\plotpoint}}
\put(316,451){\usebox{\plotpoint}}
\put(317,453){\usebox{\plotpoint}}
\put(318,455){\usebox{\plotpoint}}
\put(319,458){\usebox{\plotpoint}}
\put(320,460){\usebox{\plotpoint}}
\put(321,462){\usebox{\plotpoint}}
\put(322,465){\usebox{\plotpoint}}
\put(323,467){\usebox{\plotpoint}}
\put(324,470){\usebox{\plotpoint}}
\put(325,472){\usebox{\plotpoint}}
\put(326,475){\usebox{\plotpoint}}
\put(327,477){\usebox{\plotpoint}}
\put(328,480){\usebox{\plotpoint}}
\put(329,483){\usebox{\plotpoint}}
\put(330,486){\usebox{\plotpoint}}
\put(331,490){\usebox{\plotpoint}}
\put(332,493){\usebox{\plotpoint}}
\put(333,496){\usebox{\plotpoint}}
\put(334,500){\usebox{\plotpoint}}
\put(335,503){\usebox{\plotpoint}}
\put(336,506){\usebox{\plotpoint}}
\put(337,509){\usebox{\plotpoint}}
\put(338,512){\usebox{\plotpoint}}
\put(339,515){\usebox{\plotpoint}}
\put(340,518){\usebox{\plotpoint}}
\put(341,521){\usebox{\plotpoint}}
\put(342,524){\usebox{\plotpoint}}
\put(343,528){\usebox{\plotpoint}}
\put(344,531){\usebox{\plotpoint}}
\put(345,534){\usebox{\plotpoint}}
\put(346,538){\usebox{\plotpoint}}
\put(347,542){\usebox{\plotpoint}}
\put(348,546){\usebox{\plotpoint}}
\put(349,550){\usebox{\plotpoint}}
\put(350,554){\usebox{\plotpoint}}
\put(351,557){\rule[-0.500pt]{1.000pt}{1.253pt}}
\put(352,563){\rule[-0.500pt]{1.000pt}{1.253pt}}
\put(353,568){\rule[-0.500pt]{1.000pt}{1.253pt}}
\put(354,573){\rule[-0.500pt]{1.000pt}{1.253pt}}
\put(355,578){\rule[-0.500pt]{1.000pt}{1.253pt}}
\put(356,584){\rule[-0.500pt]{1.000pt}{1.164pt}}
\put(357,588){\rule[-0.500pt]{1.000pt}{1.164pt}}
\put(358,593){\rule[-0.500pt]{1.000pt}{1.164pt}}
\put(359,598){\rule[-0.500pt]{1.000pt}{1.164pt}}
\put(360,603){\rule[-0.500pt]{1.000pt}{1.164pt}}
\put(361,608){\rule[-0.500pt]{1.000pt}{1.164pt}}
\put(362,612){\rule[-0.500pt]{1.000pt}{1.365pt}}
\put(363,618){\rule[-0.500pt]{1.000pt}{1.365pt}}
\put(364,624){\rule[-0.500pt]{1.000pt}{1.365pt}}
\put(365,630){\rule[-0.500pt]{1.000pt}{1.365pt}}
\put(366,635){\rule[-0.500pt]{1.000pt}{1.365pt}}
\put(367,641){\rule[-0.500pt]{1.000pt}{1.365pt}}
\put(368,647){\rule[-0.500pt]{1.000pt}{1.566pt}}
\put(369,653){\rule[-0.500pt]{1.000pt}{1.566pt}}
\put(370,660){\rule[-0.500pt]{1.000pt}{1.566pt}}
\put(371,666){\rule[-0.500pt]{1.000pt}{1.566pt}}
\put(372,673){\rule[-0.500pt]{1.000pt}{1.566pt}}
\put(373,679){\rule[-0.500pt]{1.000pt}{1.566pt}}
\put(374,686){\rule[-0.500pt]{1.000pt}{1.847pt}}
\put(375,693){\rule[-0.500pt]{1.000pt}{1.847pt}}
\put(376,701){\rule[-0.500pt]{1.000pt}{1.847pt}}
\put(377,709){\rule[-0.500pt]{1.000pt}{1.847pt}}
\put(378,716){\rule[-0.500pt]{1.000pt}{1.847pt}}
\put(379,724){\rule[-0.500pt]{1.000pt}{1.847pt}}
\put(380,732){\rule[-0.500pt]{1.000pt}{2.602pt}}
\put(381,742){\rule[-0.500pt]{1.000pt}{2.602pt}}
\put(382,753){\rule[-0.500pt]{1.000pt}{2.602pt}}
\put(383,764){\rule[-0.500pt]{1.000pt}{2.602pt}}
\put(384,775){\rule[-0.500pt]{1.000pt}{2.602pt}}
\put(385,785){\usebox{\plotpoint}}
\put(264,368){\usebox{\plotpoint}}
\put(264,366){\usebox{\plotpoint}}
\put(265,365){\usebox{\plotpoint}}
\put(266,364){\usebox{\plotpoint}}
\put(267,363){\usebox{\plotpoint}}
\put(268,362){\usebox{\plotpoint}}
\put(269,361){\usebox{\plotpoint}}
\put(270,361){\usebox{\plotpoint}}
\put(270,361){\usebox{\plotpoint}}
\put(271,360){\usebox{\plotpoint}}
\put(272,359){\usebox{\plotpoint}}
\put(273,358){\usebox{\plotpoint}}
\put(274,357){\usebox{\plotpoint}}
\put(275,356){\usebox{\plotpoint}}
\put(276,353){\usebox{\plotpoint}}
\put(277,352){\usebox{\plotpoint}}
\put(278,351){\usebox{\plotpoint}}
\put(279,350){\usebox{\plotpoint}}
\put(280,349){\usebox{\plotpoint}}
\put(281,349){\usebox{\plotpoint}}
\put(282,348){\usebox{\plotpoint}}
\put(283,347){\usebox{\plotpoint}}
\put(284,346){\usebox{\plotpoint}}
\put(285,345){\usebox{\plotpoint}}
\put(287,344){\usebox{\plotpoint}}
\put(288,343){\usebox{\plotpoint}}
\put(289,342){\usebox{\plotpoint}}
\put(290,341){\usebox{\plotpoint}}
\put(291,340){\usebox{\plotpoint}}
\put(293,339){\usebox{\plotpoint}}
\put(294,338){\usebox{\plotpoint}}
\put(295,337){\usebox{\plotpoint}}
\put(296,336){\usebox{\plotpoint}}
\put(297,335){\usebox{\plotpoint}}
\put(299,334){\usebox{\plotpoint}}
\put(300,333){\usebox{\plotpoint}}
\put(301,332){\usebox{\plotpoint}}
\put(302,331){\usebox{\plotpoint}}
\put(304,330){\usebox{\plotpoint}}
\put(305,329){\usebox{\plotpoint}}
\put(306,328){\usebox{\plotpoint}}
\put(307,327){\usebox{\plotpoint}}
\put(308,326){\usebox{\plotpoint}}
\put(310,325){\usebox{\plotpoint}}
\put(311,324){\usebox{\plotpoint}}
\put(313,323){\usebox{\plotpoint}}
\put(314,322){\usebox{\plotpoint}}
\put(316,321){\usebox{\plotpoint}}
\put(317,320){\usebox{\plotpoint}}
\put(319,319){\usebox{\plotpoint}}
\put(320,318){\usebox{\plotpoint}}
\put(322,317){\usebox{\plotpoint}}
\put(323,316){\usebox{\plotpoint}}
\put(325,315){\usebox{\plotpoint}}
\put(326,314){\usebox{\plotpoint}}
\put(328,313){\usebox{\plotpoint}}
\put(329,312){\usebox{\plotpoint}}
\put(331,311){\usebox{\plotpoint}}
\put(332,310){\usebox{\plotpoint}}
\put(334,309){\usebox{\plotpoint}}
\put(336,308){\usebox{\plotpoint}}
\put(337,307){\usebox{\plotpoint}}
\put(339,306){\usebox{\plotpoint}}
\put(341,305){\usebox{\plotpoint}}
\put(343,304){\usebox{\plotpoint}}
\put(345,303){\usebox{\plotpoint}}
\put(347,302){\usebox{\plotpoint}}
\put(349,301){\usebox{\plotpoint}}
\put(351,300){\usebox{\plotpoint}}
\put(352,299){\usebox{\plotpoint}}
\put(354,298){\usebox{\plotpoint}}
\put(355,297){\usebox{\plotpoint}}
\put(358,296){\usebox{\plotpoint}}
\put(360,295){\usebox{\plotpoint}}
\put(362,294){\usebox{\plotpoint}}
\put(364,293){\usebox{\plotpoint}}
\put(366,292){\usebox{\plotpoint}}
\put(368,291){\usebox{\plotpoint}}
\put(371,290){\usebox{\plotpoint}}
\put(374,289){\usebox{\plotpoint}}
\put(376,288){\usebox{\plotpoint}}
\put(378,287){\usebox{\plotpoint}}
\put(380,286){\usebox{\plotpoint}}
\put(382,285){\usebox{\plotpoint}}
\put(385,284){\usebox{\plotpoint}}
\put(387,283){\usebox{\plotpoint}}
\put(389,282){\usebox{\plotpoint}}
\put(391,281){\usebox{\plotpoint}}
\put(394,280){\usebox{\plotpoint}}
\put(397,279){\usebox{\plotpoint}}
\put(400,278){\usebox{\plotpoint}}
\put(403,277){\usebox{\plotpoint}}
\put(405,276){\usebox{\plotpoint}}
\put(408,275){\usebox{\plotpoint}}
\put(411,274){\usebox{\plotpoint}}
\put(414,273){\usebox{\plotpoint}}
\put(417,272){\usebox{\plotpoint}}
\put(420,271){\usebox{\plotpoint}}
\put(423,270){\usebox{\plotpoint}}
\put(426,269){\usebox{\plotpoint}}
\put(429,268){\usebox{\plotpoint}}
\put(432,267){\usebox{\plotpoint}}
\put(434,266){\usebox{\plotpoint}}
\put(437,265){\rule[-0.500pt]{1.445pt}{1.000pt}}
\put(443,264){\usebox{\plotpoint}}
\put(446,263){\usebox{\plotpoint}}
\put(449,262){\usebox{\plotpoint}}
\put(452,261){\usebox{\plotpoint}}
\put(455,260){\rule[-0.500pt]{1.204pt}{1.000pt}}
\put(460,259){\usebox{\plotpoint}}
\put(463,258){\usebox{\plotpoint}}
\put(466,257){\rule[-0.500pt]{1.445pt}{1.000pt}}
\put(472,256){\usebox{\plotpoint}}
\put(475,255){\usebox{\plotpoint}}
\put(478,254){\rule[-0.500pt]{1.445pt}{1.000pt}}
\put(484,253){\rule[-0.500pt]{1.204pt}{1.000pt}}
\put(489,252){\usebox{\plotpoint}}
\put(492,251){\usebox{\plotpoint}}
\put(495,250){\rule[-0.500pt]{1.445pt}{1.000pt}}
\put(501,249){\rule[-0.500pt]{1.445pt}{1.000pt}}
\put(507,248){\rule[-0.500pt]{1.204pt}{1.000pt}}
\put(512,247){\usebox{\plotpoint}}
\put(515,246){\usebox{\plotpoint}}
\put(518,245){\rule[-0.500pt]{1.445pt}{1.000pt}}
\put(524,244){\rule[-0.500pt]{1.445pt}{1.000pt}}
\put(530,243){\rule[-0.500pt]{1.445pt}{1.000pt}}
\put(536,242){\rule[-0.500pt]{1.204pt}{1.000pt}}
\put(541,241){\rule[-0.500pt]{1.445pt}{1.000pt}}
\put(547,240){\rule[-0.500pt]{1.445pt}{1.000pt}}
\put(553,239){\rule[-0.500pt]{1.445pt}{1.000pt}}
\put(559,238){\rule[-0.500pt]{1.204pt}{1.000pt}}
\put(564,237){\rule[-0.500pt]{1.445pt}{1.000pt}}
\put(570,236){\rule[-0.500pt]{1.445pt}{1.000pt}}
\put(576,235){\rule[-0.500pt]{1.445pt}{1.000pt}}
\put(582,234){\rule[-0.500pt]{1.445pt}{1.000pt}}
\put(588,233){\rule[-0.500pt]{2.650pt}{1.000pt}}
\put(599,232){\rule[-0.500pt]{1.445pt}{1.000pt}}
\put(605,231){\rule[-0.500pt]{1.445pt}{1.000pt}}
\put(611,230){\rule[-0.500pt]{1.204pt}{1.000pt}}
\put(616,229){\rule[-0.500pt]{2.891pt}{1.000pt}}
\put(628,228){\rule[-0.500pt]{1.445pt}{1.000pt}}
\put(634,227){\rule[-0.500pt]{1.445pt}{1.000pt}}
\put(640,226){\rule[-0.500pt]{2.650pt}{1.000pt}}
\put(651,225){\rule[-0.500pt]{1.445pt}{1.000pt}}
\put(657,224){\rule[-0.500pt]{2.650pt}{1.000pt}}
\put(668,223){\rule[-0.500pt]{1.445pt}{1.000pt}}
\put(674,222){\rule[-0.500pt]{2.891pt}{1.000pt}}
\put(686,221){\rule[-0.500pt]{2.650pt}{1.000pt}}
\put(697,220){\rule[-0.500pt]{1.445pt}{1.000pt}}
\put(703,219){\rule[-0.500pt]{2.891pt}{1.000pt}}
\put(715,218){\rule[-0.500pt]{2.650pt}{1.000pt}}
\put(726,217){\rule[-0.500pt]{2.891pt}{1.000pt}}
\put(738,216){\rule[-0.500pt]{2.650pt}{1.000pt}}
\put(749,215){\rule[-0.500pt]{2.891pt}{1.000pt}}
\put(761,214){\rule[-0.500pt]{2.650pt}{1.000pt}}
\put(772,213){\rule[-0.500pt]{2.891pt}{1.000pt}}
\put(784,212){\rule[-0.500pt]{2.891pt}{1.000pt}}
\put(796,211){\rule[-0.500pt]{4.095pt}{1.000pt}}
\put(813,210){\rule[-0.500pt]{2.650pt}{1.000pt}}
\put(824,209){\rule[-0.500pt]{2.891pt}{1.000pt}}
\sbox{\plotpoint}{\rule[-0.175pt]{0.350pt}{0.350pt}}%
\put(264,368){\usebox{\plotpoint}}
\put(264,368){\rule[-0.175pt]{137.795pt}{0.350pt}}
\put(516,242){\usebox{\plotpoint}}
\put(516,242){\rule[-0.175pt]{0.350pt}{4.176pt}}
\put(515,259){\rule[-0.175pt]{0.350pt}{4.176pt}}
\put(514,276){\rule[-0.175pt]{0.350pt}{4.176pt}}
\put(513,294){\rule[-0.175pt]{0.350pt}{0.482pt}}
\put(512,296){\rule[-0.175pt]{0.350pt}{0.482pt}}
\put(511,298){\rule[-0.175pt]{0.350pt}{0.482pt}}
\put(510,300){\rule[-0.175pt]{0.350pt}{0.401pt}}
\put(509,301){\rule[-0.175pt]{0.350pt}{0.401pt}}
\put(508,303){\rule[-0.175pt]{0.350pt}{0.401pt}}
\put(507,304){\rule[-0.175pt]{0.350pt}{0.402pt}}
\put(506,306){\rule[-0.175pt]{0.350pt}{0.401pt}}
\put(505,308){\rule[-0.175pt]{0.350pt}{0.401pt}}
\put(504,309){\usebox{\plotpoint}}
\put(503,311){\usebox{\plotpoint}}
\put(502,312){\usebox{\plotpoint}}
\put(501,314){\usebox{\plotpoint}}
\put(500,315){\usebox{\plotpoint}}
\put(499,316){\rule[-0.175pt]{0.350pt}{0.361pt}}
\put(498,318){\rule[-0.175pt]{0.350pt}{0.361pt}}
\put(497,320){\rule[-0.175pt]{0.350pt}{0.361pt}}
\put(496,321){\rule[-0.175pt]{0.350pt}{0.361pt}}
\put(495,323){\rule[-0.175pt]{0.350pt}{0.361pt}}
\put(494,324){\rule[-0.175pt]{0.350pt}{0.361pt}}
\put(487,326){\rule[-0.175pt]{1.373pt}{0.350pt}}
\put(481,327){\rule[-0.175pt]{1.373pt}{0.350pt}}
\put(475,328){\rule[-0.175pt]{1.373pt}{0.350pt}}
\put(470,329){\rule[-0.175pt]{1.373pt}{0.350pt}}
\put(464,330){\rule[-0.175pt]{1.373pt}{0.350pt}}
\put(458,331){\rule[-0.175pt]{1.373pt}{0.350pt}}
\put(453,332){\rule[-0.175pt]{1.373pt}{0.350pt}}
\put(447,333){\rule[-0.175pt]{1.373pt}{0.350pt}}
\put(441,334){\rule[-0.175pt]{1.373pt}{0.350pt}}
\put(436,335){\rule[-0.175pt]{1.373pt}{0.350pt}}
\put(432,336){\rule[-0.175pt]{0.873pt}{0.350pt}}
\put(428,337){\rule[-0.175pt]{0.873pt}{0.350pt}}
\put(425,338){\rule[-0.175pt]{0.873pt}{0.350pt}}
\put(421,339){\rule[-0.175pt]{0.873pt}{0.350pt}}
\put(417,340){\rule[-0.175pt]{0.873pt}{0.350pt}}
\put(414,341){\rule[-0.175pt]{0.873pt}{0.350pt}}
\put(410,342){\rule[-0.175pt]{0.873pt}{0.350pt}}
\put(407,343){\rule[-0.175pt]{0.873pt}{0.350pt}}
\put(403,344){\rule[-0.175pt]{0.873pt}{0.350pt}}
\put(399,345){\rule[-0.175pt]{0.873pt}{0.350pt}}
\put(396,346){\rule[-0.175pt]{0.873pt}{0.350pt}}
\put(392,347){\rule[-0.175pt]{0.873pt}{0.350pt}}
\put(388,348){\rule[-0.175pt]{0.873pt}{0.350pt}}
\put(385,349){\rule[-0.175pt]{0.873pt}{0.350pt}}
\put(381,350){\rule[-0.175pt]{0.873pt}{0.350pt}}
\put(378,351){\rule[-0.175pt]{0.873pt}{0.350pt}}
\put(372,352){\rule[-0.175pt]{1.373pt}{0.350pt}}
\put(366,353){\rule[-0.175pt]{1.373pt}{0.350pt}}
\put(360,354){\rule[-0.175pt]{1.373pt}{0.350pt}}
\put(355,355){\rule[-0.175pt]{1.373pt}{0.350pt}}
\put(349,356){\rule[-0.175pt]{1.373pt}{0.350pt}}
\put(343,357){\rule[-0.175pt]{1.373pt}{0.350pt}}
\put(338,358){\rule[-0.175pt]{1.373pt}{0.350pt}}
\put(332,359){\rule[-0.175pt]{1.373pt}{0.350pt}}
\put(326,360){\rule[-0.175pt]{1.373pt}{0.350pt}}
\put(321,361){\rule[-0.175pt]{1.373pt}{0.350pt}}
\put(311,362){\rule[-0.175pt]{2.289pt}{0.350pt}}
\put(302,363){\rule[-0.175pt]{2.289pt}{0.350pt}}
\put(292,364){\rule[-0.175pt]{2.289pt}{0.350pt}}
\put(283,365){\rule[-0.175pt]{2.289pt}{0.350pt}}
\put(273,366){\rule[-0.175pt]{2.289pt}{0.350pt}}
\put(264,367){\rule[-0.175pt]{2.289pt}{0.350pt}}
\put(366,630){\usebox{\plotpoint}}
\put(366,609){\rule[-0.175pt]{0.350pt}{5.059pt}}
\put(365,588){\rule[-0.175pt]{0.350pt}{5.059pt}}
\put(364,567){\rule[-0.175pt]{0.350pt}{5.059pt}}
\put(363,546){\rule[-0.175pt]{0.350pt}{5.059pt}}
\put(362,525){\rule[-0.175pt]{0.350pt}{5.059pt}}
\put(361,514){\rule[-0.175pt]{0.350pt}{2.505pt}}
\put(360,504){\rule[-0.175pt]{0.350pt}{2.505pt}}
\put(359,493){\rule[-0.175pt]{0.350pt}{2.505pt}}
\put(358,483){\rule[-0.175pt]{0.350pt}{2.505pt}}
\put(357,473){\rule[-0.175pt]{0.350pt}{2.505pt}}
\put(356,467){\rule[-0.175pt]{0.350pt}{1.325pt}}
\put(355,462){\rule[-0.175pt]{0.350pt}{1.325pt}}
\put(354,452){\rule[-0.175pt]{0.350pt}{2.409pt}}
\put(353,446){\rule[-0.175pt]{0.350pt}{1.325pt}}
\put(352,441){\rule[-0.175pt]{0.350pt}{1.325pt}}
\put(351,436){\rule[-0.175pt]{0.350pt}{1.204pt}}
\put(350,431){\rule[-0.175pt]{0.350pt}{1.204pt}}
\put(349,420){\rule[-0.175pt]{0.350pt}{2.650pt}}
\put(348,412){\rule[-0.175pt]{0.350pt}{1.927pt}}
\put(347,404){\rule[-0.175pt]{0.350pt}{1.927pt}}
\put(346,394){\rule[-0.175pt]{0.350pt}{2.409pt}}
\put(345,391){\rule[-0.175pt]{0.350pt}{0.602pt}}
\put(344,389){\rule[-0.175pt]{0.350pt}{0.602pt}}
\put(343,387){\usebox{\plotpoint}}
\put(342,386){\usebox{\plotpoint}}
\put(341,385){\usebox{\plotpoint}}
\put(340,384){\usebox{\plotpoint}}
\put(339,383){\usebox{\plotpoint}}
\put(336,383){\usebox{\plotpoint}}
\put(335,382){\usebox{\plotpoint}}
\put(334,381){\usebox{\plotpoint}}
\put(333,380){\usebox{\plotpoint}}
\put(331,379){\rule[-0.175pt]{0.361pt}{0.350pt}}
\put(330,378){\rule[-0.175pt]{0.361pt}{0.350pt}}
\put(328,377){\rule[-0.175pt]{0.361pt}{0.350pt}}
\put(327,376){\rule[-0.175pt]{0.361pt}{0.350pt}}
\put(324,375){\rule[-0.175pt]{0.723pt}{0.350pt}}
\put(321,374){\rule[-0.175pt]{0.723pt}{0.350pt}}
\put(309,373){\rule[-0.175pt]{2.746pt}{0.350pt}}
\put(298,372){\rule[-0.175pt]{2.746pt}{0.350pt}}
\put(286,371){\rule[-0.175pt]{2.746pt}{0.350pt}}
\put(275,370){\rule[-0.175pt]{2.746pt}{0.350pt}}
\put(264,369){\rule[-0.175pt]{2.746pt}{0.350pt}}
\put(264,368){\usebox{\plotpoint}}
\end{picture}
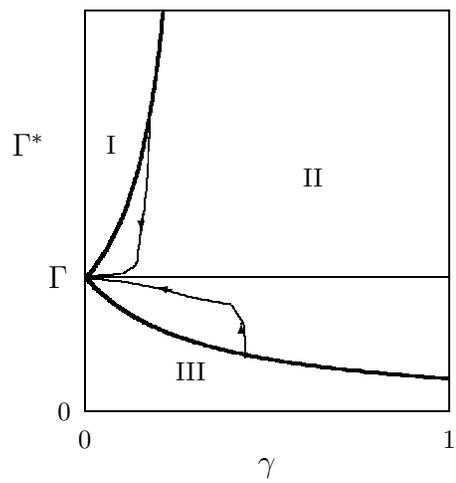
\end{document}